\documentclass{article}

\usepackage{arxiv}

\usepackage[utf8]{inputenc} 
\usepackage[T1]{fontenc}    
\usepackage{hyperref}       
\usepackage{url}            
\usepackage{booktabs}       
\usepackage{amsfonts}       
\usepackage{nicefrac}       
\usepackage{microtype}      
\usepackage{lipsum}
\usepackage{tikz}
\usepackage{amssymb}
\usepackage{amsmath}
\usepackage{float}
\usepackage{subcaption}

\title{Forecasting COVID-19 Pandemic in Mozambique and Estimating Possible Scenarios}

\author{
  Cláudio Moisés Paulo\\
  Grupo de Astrofísica e Ciências Espaciais\\
  Universidade Eduardo Mondlane\\
  Av. Julius Nyerere, nº 3453,Campus Principal\\
  \texttt{claudiompaulo@uem.mz}\\
   \And
  Felipe Nunes Fontinele\\
  Department of Physics, Faculty of Sciences\\
  University of Alberta\\
  4-181 CCIS, Edmonton, Alberta, Canada T6G 2E1\\
  \texttt{feradofogo@hotmail.com} \\
   \And
  Pedro Henrique P. Cintra\\
  Instituto de Física\\
  Universidade de Brasília\\
  Campus Darcy Ribeiro, Asa norte, Brasília, Brazil\\
  \texttt{pedrohpc96@hotmail.com} \\
}

\begin{document}
\maketitle

\begin{abstract}
COVID-19 is now the largest pandemic crisis of this century, with over 16 million registered cases worldwide. African countries have now begun registering an increasing number of cases, yet, not many models developed focus in specific African countries. In our study we use a simple SEIR model to evaluate and predict future scenarios regarding the pandemic crisis in Mozambique. We compare the effect of different policies on the infection curve and estimate epidemiological parameters such as the current infection reproduction number $\mathcal{R}_t$ and the growth rate $g$. We have found a low value for $\mathcal{R}_t$, ranging from 1.11 to 1.48 and a positive growth rate, between $g$ = 0.22 to 0.27. Our simulations also suggest that a lockdown shows potential for reducing the infection peak height in 28\%, on average, ranging from 20 to 36\%.
\end{abstract}

\keywords{SEIR \and Mozambique \and COVID-19 \and Forecast \and Coronavirus}

\section{Introduction}

The ongoing Coronavirus disease 2019 (COVID-19) pandemic crisis has now over 16 million registered cases worldwide and more than 600 thousand deaths, according to the situation report 186 by the World Health Organization (\url{https://www.who.int/emergencies/diseases/novel-coronavirus-2019/situation-reports}). COVID-19 is a respiratory disease caused by the SARS-CoV-2 virus, novel species of the \textit{betacoronavirus} genus, it shows high affinity with Angiotensin converting enzyme 2 (ACE2) receptor in human cells \cite{lan2020structure, wang2020structural, yan2020structural}. The most probable origin of the virus was traced back to bat populations by genetic studies \cite{andersen2020proximal, tang2020origin}, with the Pangolin being a probable intermediate host \cite{lam2020identifying}.

Nowadays, the disease presents a growing behavior throughout the African continent. With no recent scientific studies concerning the projection of the pandemic dynamics for Mozambique, local authorities must rely on international measures in order to control the spreading rate of COVID-19. This study aims at accessing the problem in Mozambique, estimating important epidemiological factors such as the growth rate $g$ and the current value of the infection reproduction number $\mathcal{R}_t$. Using a simple SEIR model including the dead compartment, we rely on daily epidemiological information available by the government (\url{https://covid19.ins.gov.mz/documentos/}); concerning the cumulative number of infections until July 24, cumulative number of recoveries and deaths, to forecast the possible futures the pandemic might take on the country.

Mathematical models are now worldwide used for forecasting \cite{fernandez2020estimating, iwata2020simulation, shoukat2020projecting, walker2020global}, data analysis \cite{dehning2020inferring} and economic impact \cite{toda2020susceptible, atkeson2020will} regarding the current pandemic crisis. SIR and SEIR models are the most simple and used ones for evaluation of the outbreak dynamics \cite{anastassopoulou2020data}. Although more complex models are frequently employed and might yield more specific results; such as models concerning quarantine, asymptomatic and hospitalizations, these models make use of many unknown parameters, presenting a challenge when trying to copy reality in simulations.

\section{Model Description}

We apply a modified SEIR model to include deaths as a new compartment. The model relies on the following chain of events: A susceptible individual from a population $N$ of constant size is exposed to the virus, joining the exposed group, in which the person does not show symptoms of the disease yet, after some time (incubation period) this exposed person joins the infected group, where the symptoms onset. From this stage the patient either recovers or dies.

\begin{figure}[ht]
    \centering
    \includegraphics[width=0.7\textwidth]{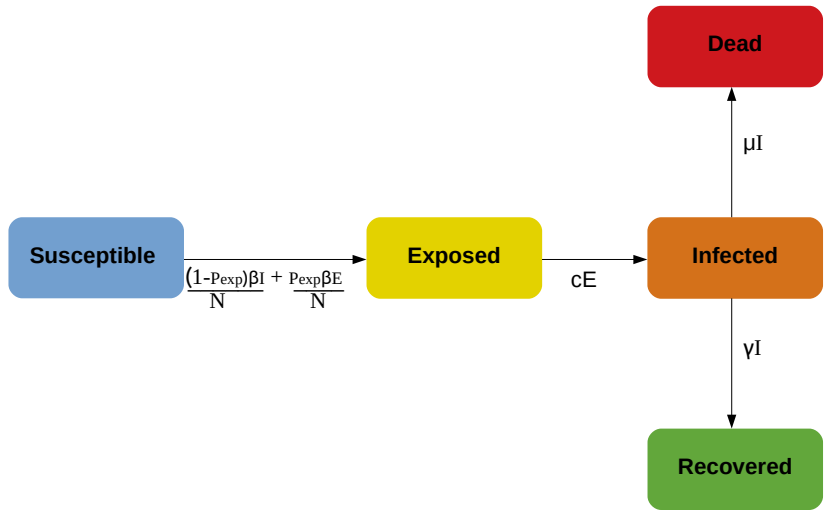}
    \caption{Representation of the SEIRD model described above.}
    \label{fig.SEIRD}
\end{figure}

The model treats of populations, therefore these transitions between groups are given by rates of changes in a specific population. Susceptible individuals, denoted by $S$, become infected through contact with infected ($I$) or exposed ($E$) individuals at a rate proportional to the density of infected and exposed $[(1-P_{exp})\beta I + P_{exp} \beta E]/N$, where $P_{exp}$ is the percentage of infections caused by the exposed population and $\beta$ is the infection rate. The exposed population declines as patients become infected by a rate $c$ proportional to the inversion of the incubation period $c = 1/\tau$, with $\tau$ being the incubation period. Once inside the infected population, patients now transit to recovered ($R$) or dead ($D$) groups through constant rates $\gamma$ and $\mu$ respectively. These rates are proportional to the infection fatality rate (IFR) and the average time taken from symptoms onset to recovery $\tau_r$ and the average time taken from symptoms onset to death $\tau_d$ (Figure \ref{fig.SEIRD}).

The dynamics described by these populations is mathematically represented by the following set of differential equations

\begin{align}
\label{eq.1}
    &\frac{dS}{dt} = - \frac{(1-P_{exp})\beta}{N} I S - \frac{P_{exp}\beta}{N} E S \\
\label{eq.2}
    &\frac{dE}{dt} = \frac{(1-P_{exp})\beta}{N} I S + \frac{P_{exp}\beta}{N} E - c E \\
\label{eq.3}
    &\frac{dI}{dt} = c E - (\gamma + \mu) I \\
\label{eq.4}
    &\frac{dR}{dt} = \gamma I \\
\label{eq.5}
    &\frac{dD}{dt} = \mu I
\end{align}

\noindent where $\gamma$ and $\mu$ are given by

\begin{align}
\label{eq.6}
    &\gamma = \frac{(1-IFR)}{\tau_r} \\
\nonumber
    &\mu = \frac{IFR}{\tau_d}.
\end{align}

These equations are subjected to a set of initial conditions at $t=0$; here we assume $S(t=0) = S_0 \approx N$, $E(t=0) = E_0$, $I(t=0) = I_0$, $R(t=0) = 0$ and $D(t=0) = 0$. An important consideration taken into account by the model is that the total population $N$ is constant, which can be noted by summing equations \eqref{eq.1} - \eqref{eq.5} and noticing that it equals to 0, therefore $dN/dt = dS/dt + dE/dt + dI/dt + dR/dt + dD/dt = 0 \Rightarrow N = \text{constant}$. From this type of model we retrieve the basic reproduction number, from which we calculate the current infection reproduction number using the current value for $\beta$ instead of the natural value without intervention policies, for the SEIRD model which was already derived in a previous work \cite{cintra2020mathematical}, where was made use of a method developed in \cite{van2002reproduction}, resulting in

\begin{align}
\label{eq.7}
    \mathcal{R}_0 = \frac{P_{exp} \beta\left[\gamma + (1-P_{exp})\beta\right] + (1-P_{exp})\beta c}{c(\gamma + \mu)}
\end{align}

Another important quantity derived from the model and widely used to access the growth of the outbreak in a particular region is the growth rate, given by

\begin{align}
\label{eq.8}
    g = \beta - (\gamma + \mu)
\end{align}

\noindent which states the difference between new infections and new outcomes from previous infections.

\subsection{Non-Pharmaceutical Intervention}

The next issue treated by our model is the description of non-pharmaceutical interventions such as social isolation and lockdown. We assume these actions alter the infection rate as time progresses; the core of the infection rate $\beta$ embeds the average daily contacts an individual has and the probability of being infected at each contact. Therefore, we consider a time-variant $\beta$ obeying a logistic behavior described by the following equation

\begin{align}
\label{eq.9}
    \frac{\beta(t)}{N} = \frac{P_{red}\beta_i/N}{1+\Gamma e^{b(t-t_c)}} + (1-P_{red})\beta_i/N
\end{align}

\noindent where $P_{red}$ is the percentage of reduction suffered by $\beta(t)$ as a result from the intervention, $\beta_i$ is the initial value $\beta(t)$, that is, the value $\beta(t)$ would acquire without the intervention. Parameters $t_c$, $b$ and $\Gamma$ are, respectively, the critical time when the intervention starts, the control on how fast $\beta$ decreases to the new value and a scale factor adjusted so the decrease of $\beta(t)$ starts at the given date; here we choose $\Gamma = 0.01$ and $b = 1$, with those parameters set the time taken for $\beta(t)$ do decrease the desirable amount represented by $P_{red}$ would be approximately 10 days.

Considering this behavior we may represent the action of different non-pharmaceutical interventions based on their effectiveness and starting date. We might also add more logistic functions with different parameters and different sign for $b$, in such a way to predict the ending of a given intervention policy.

\section{Simulations}
\label{sec.Simulations}

In order to simulate future scenarios and the actual tendency of Mozambique regarding the dynamics of the outbreak, we used a Python code freely available at \url{https://github.com/PedroHPCintra/COVID-Mozambique}. Using the non-linear least squares method, we fitted infected data constructed from the subtraction of the cumulative number by the recoveries and deaths ($I = CT - R - D$, where $CT$ is the cumulative total), from the fit we acquired values of epidemiological parameter, allowing the acquisition of the growth rate $g$ and the infection reproduction number $\mathcal{R}_t$. After fitting the data, we used the fitted parameters to simulate the future tendency of behavior, as well as future scenarios developed in section \ref{sec.Scenarios}.

\begin{figure}[H]
    \centering
    \includegraphics[width=0.6\textwidth]{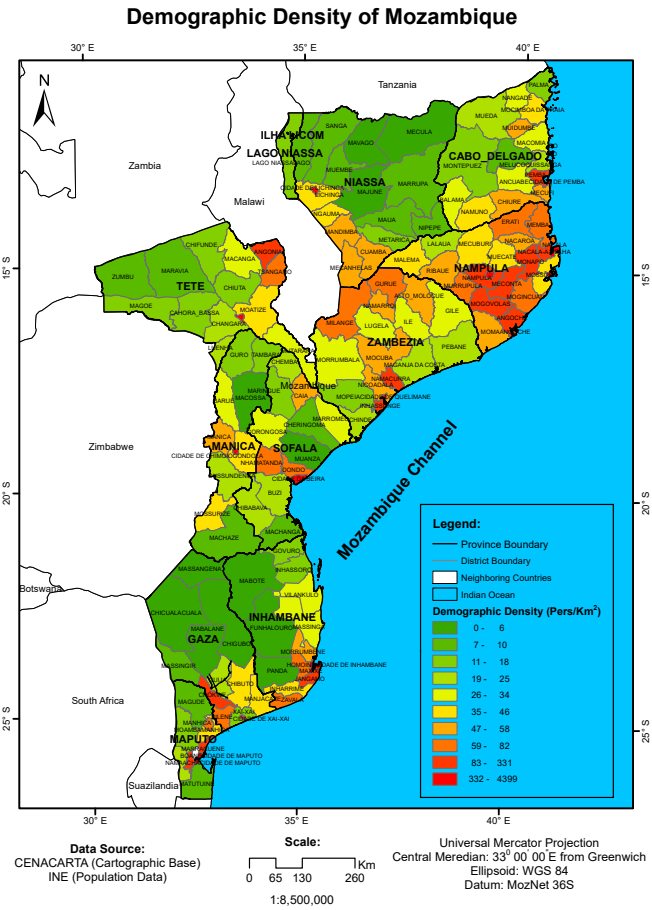}
    \caption{Demographic density of Mozambique. From the map it is possible to verify the higher demographic concentration in the region of Nampula and Zambezia, as well as localized districts of higher population per $Km^2$.}
    \label{fig.Demografia}
\end{figure}

For the fitting we decided to leave the parameters $\beta$, $E_0$ and $I_0$ free, while $\gamma$, $\mu$, $c$, $t_c$, $P_{red}$ are given in table \ref{tab.simulation}. The major problem with fitting data at initial stages of the pandemic is the computational estimation of $S_0$, which in our case is approximated to the total population $N$. The mathematical model presumes a homogeneously distributed population throughout the country, which is not a considerable reality, as shown in figure \ref{fig.Demografia}, thus, we cannot consider $N$ as being the total population of Mozambique; instead, $N$ plays the role of an effective population, which is the equivalent population of the country if the spatial distribution of people was indeed homogeneous.

Aiming to solve this limitation, we ran several simulations using different fixed values of $N$, ranging from $0.01\%$ of the population of Cabo Delgado + Maputo + Nampula to $10\%$. The choice of these regions to represent the effective $N$ of Mozambique lies in the fact that they represent the vast majority of cases, that way, we approximate the population of Mozambique to be fractions of the populations of Nampula, Cabo Delgado and Maputo. At each stage, the $\chi^2$ of the fit was taken into account and, if it was lower than 0.98, the fit was discarded. In this way, we narrow the range of possible $N$s that might describe this first pandemic wave. We discuss in more detail how $N$ changes during the pandemic in section \ref{sec.Discussion}.

\begin{table}[H]
    \centering
    \begin{tabular}{|c|c|c|}
    \hline
    Parameter & Value & Reference \\
    \hline
    $IFR$ & 0.5\% & \cite{salje2020estimating} \\
    \hline
    $c$ & 5.1 days$^{-1}$ & \cite{li2020early, backer2020incubation, linton2020incubation, guan2020clinical} \\
    \hline
    $\tau_r$ & 10 days & \cite{bernheim2020chest} \\
    \hline
    $\tau_d$ & 18 days & \cite{ruan2020clinical} \\
    \hline
    $t_c$ & 8 days & Local government decree \\
    \hline
    $P_{red}$ & 50\% & \cite{hsiang2020effect} \\
    \hline
    $P_{exp}$ & 44\% & \cite{he2020temporal} \\
    \hline
    \end{tabular}
    \caption{Parameters used for the simulation and the fitting of epidemiological data from Mozambique.}
    \label{tab.simulation}
\end{table}

Later, when simulating the possible scenarios in section \ref{sec.Scenarios}, we added a second logistic drop to $\beta(t)$ starting at the supposed date for a lockdown imposition, at the beginning of August. The drop decreased $\beta_i$ 20\% more, that way the total decrease suffered by $\beta_i$ would be 70\%, which is consistent with the efficacy of decrease lockdowns acquire, supposing that the decrease in $\mathcal{R}_t$ \footnote{$\mathcal{R}_t$ represents the change $\mathcal{R}_0$ suffers due to intervention policies at the outbreak. While $\mathcal{R}_0$ is specific for each disease and represents the spread without any intervention, $\mathcal{R}_t$ changes with time due to intervention policies.} or in the growth rate $g$ is due to a decrease in the infection rate $\beta$ \cite{salje2020estimating}. Then, to simulate faster or slower re-openings, we considered the scenarios where the infection rate grows as fast as it decreased (fast re-opening) and the scenario where the infection rate grows 10 times slower than it decreased (slow re-opening); that is done by changing the $b$ parameter in a third logistic function with inverted sign at $b$, representing a increase, instead of a drop.

\begin{figure}[ht]
    \centering
    \includegraphics[width=0.7\textwidth]{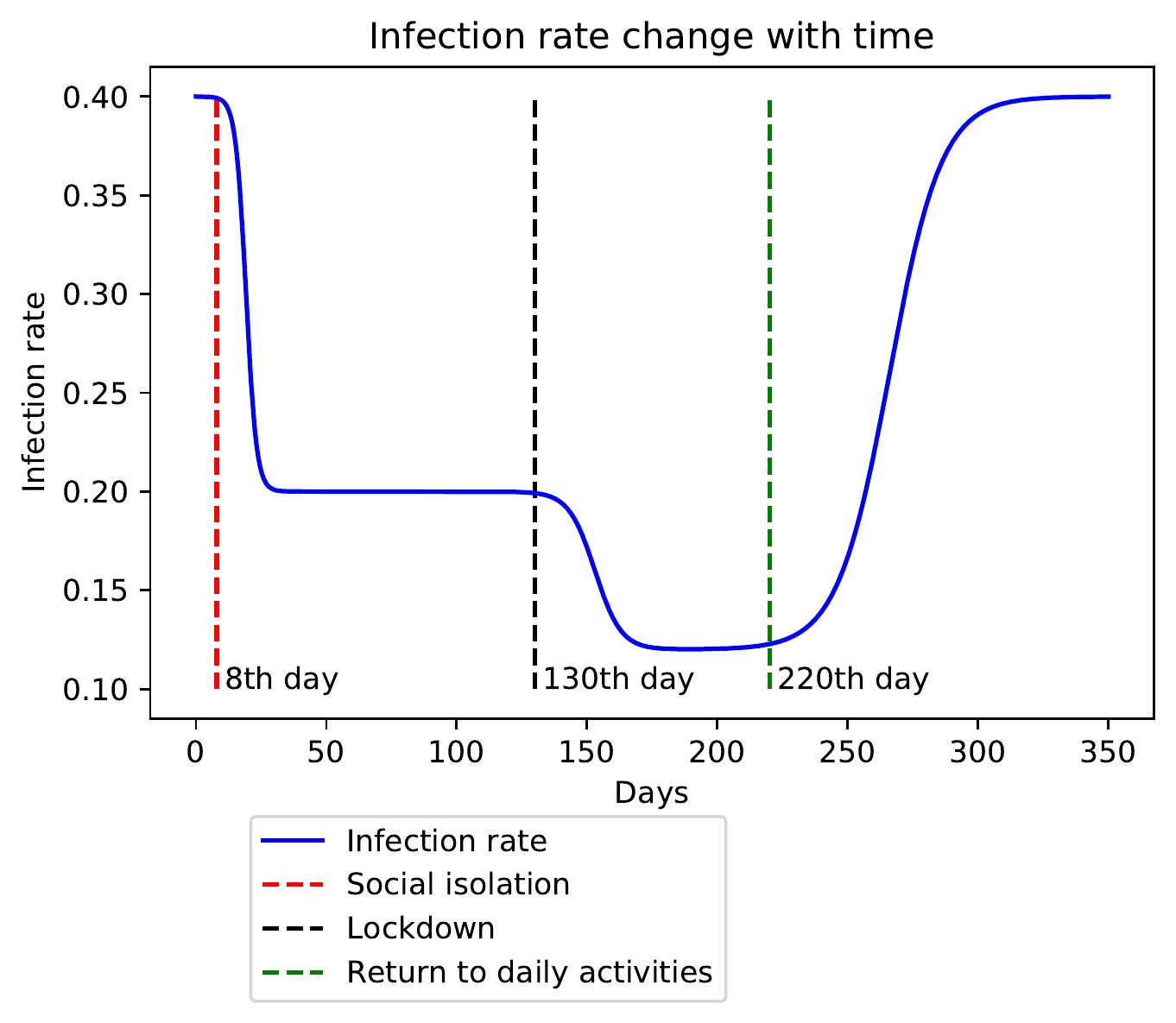}
    \caption{Representation of the temporal change in $\beta(t)$ due to intervention policies. Here we use a sum of three logistic curves with $t_c$s corresponding to the date in which each policy is applied. The $b$ parameter for the third logistic function, representing the return to daily activities, was chosen to be -0.1, in contrast to $b = 1$ at the first two logistic curves.}
    \label{fig.beta}
\end{figure}

\section{Results}
\label{sec.Results}

When performing the fitting, the minimum value of $N$ acceptable by the $\chi^2$ elimination was $N_{min} = 0.09\%$ of the population, while the largest value was $N_{max} = 0.5\%$ of the population. We then performed fittings considering $N = 0.09\%$, $0.1\%$,  $0.2\%$, $0.3\%$, $0.4\%$ and $0.5\%$ and $\beta_i = 0.4$, $I_0 = 1$ and $E_0 = 5$ as a initial guess; $\beta_i$ was chosen according to the observed value when fitting data from other countries with reliable data \cite{cintra2020estimative}. Our first adjustment of the current data yielded $\beta_i = 0.31 \pm 0.02$, $I_0 = 2 \pm 1$ and $E_0 = 7 \pm 5$.

Having $\beta_i$ at hand, we calculated the range of $g$ and $\mathcal{R}_t$ in Mozambique, using the known international values for $\gamma$, $\mu$, $c$ and $P_{exp}$ given by table \ref{tab.simulation}.

\begin{align}
\label{eq.10}
    &g = 0.25 \pm 0.03\\
\label{eq.11}
    &\mathcal{R}_t = 1.29 \pm 0.19
\end{align}

\begin{figure}[ht]
    \centering
    \includegraphics[width=0.8\textwidth]{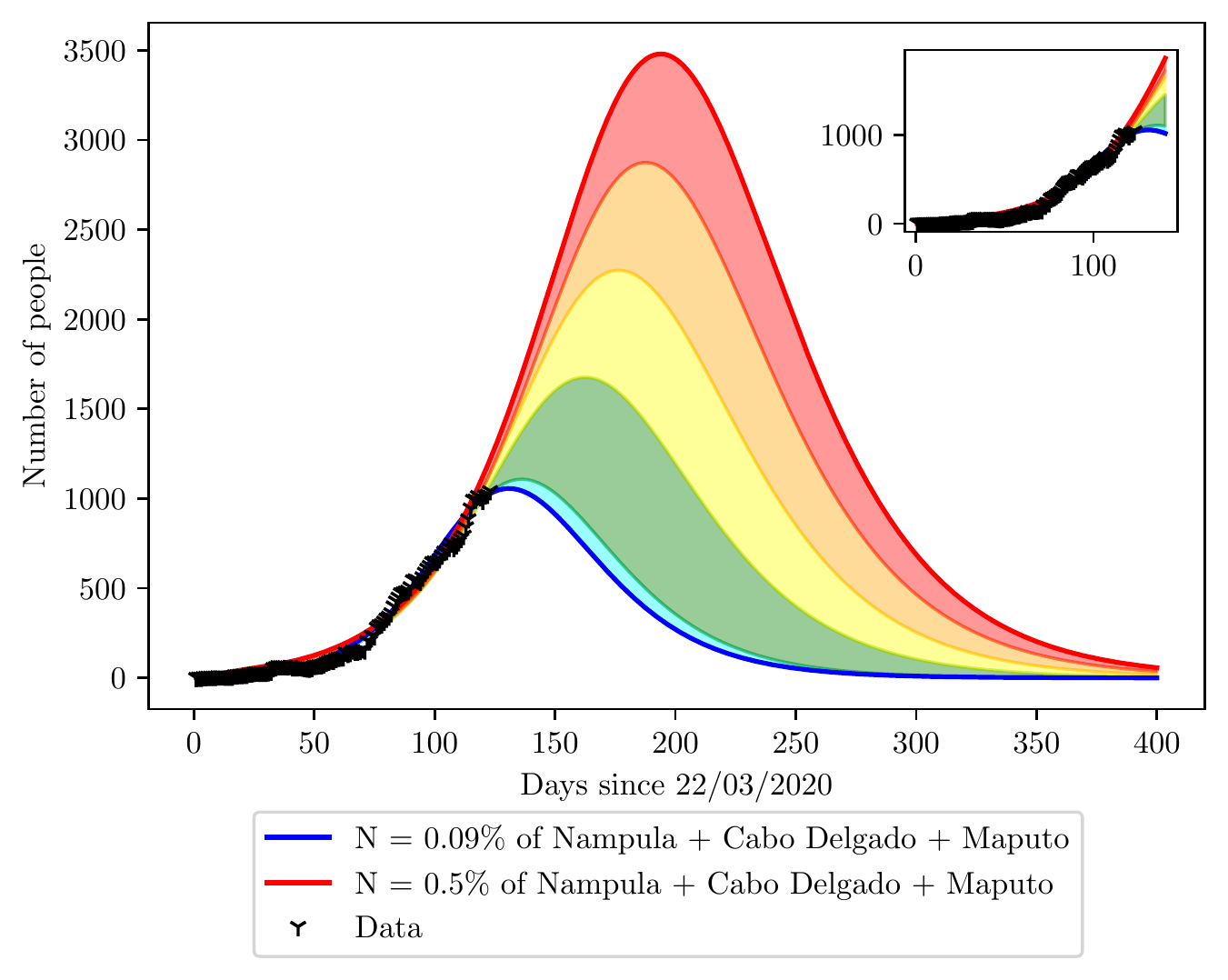}
    \caption{Predictions of possible behaviors of the outbreak taking into account the current data available. The color gradient represents each $N$ simulated, blue curve is the simulated scenario with $N = 0.09\%$, cyan region represents $N=0.1\%$, green $N=0.2\%$, yellow = $N=0.3\%$ and so on. Infection peak is estimated to be between the final of August and September, with a infection peak height between 1500 infections (on the best scenario) and 3800 (on the worse scenario)}
    \label{fig.Current}
\end{figure}

The results of the possible tendencies for the future behavior of the pandemic, according to the recent data, is shown at figure \ref{fig.Current}.

The current scenario predicted to Mozambique shows that the country might be close to an infection peak, given the current tendency; however it is also possible for the infection curve to increase at least for two more months and reach an infection peak close to 3500 cases, with a cumulative number of cases around 30000 after 400 days of pandemic.

\subsection{Future Scenarios}
\label{sec.Scenarios}

\begin{figure}[ht]
  \centering
  \begin{minipage}[b]{0.8\textwidth}
    \includegraphics[width=\textwidth]{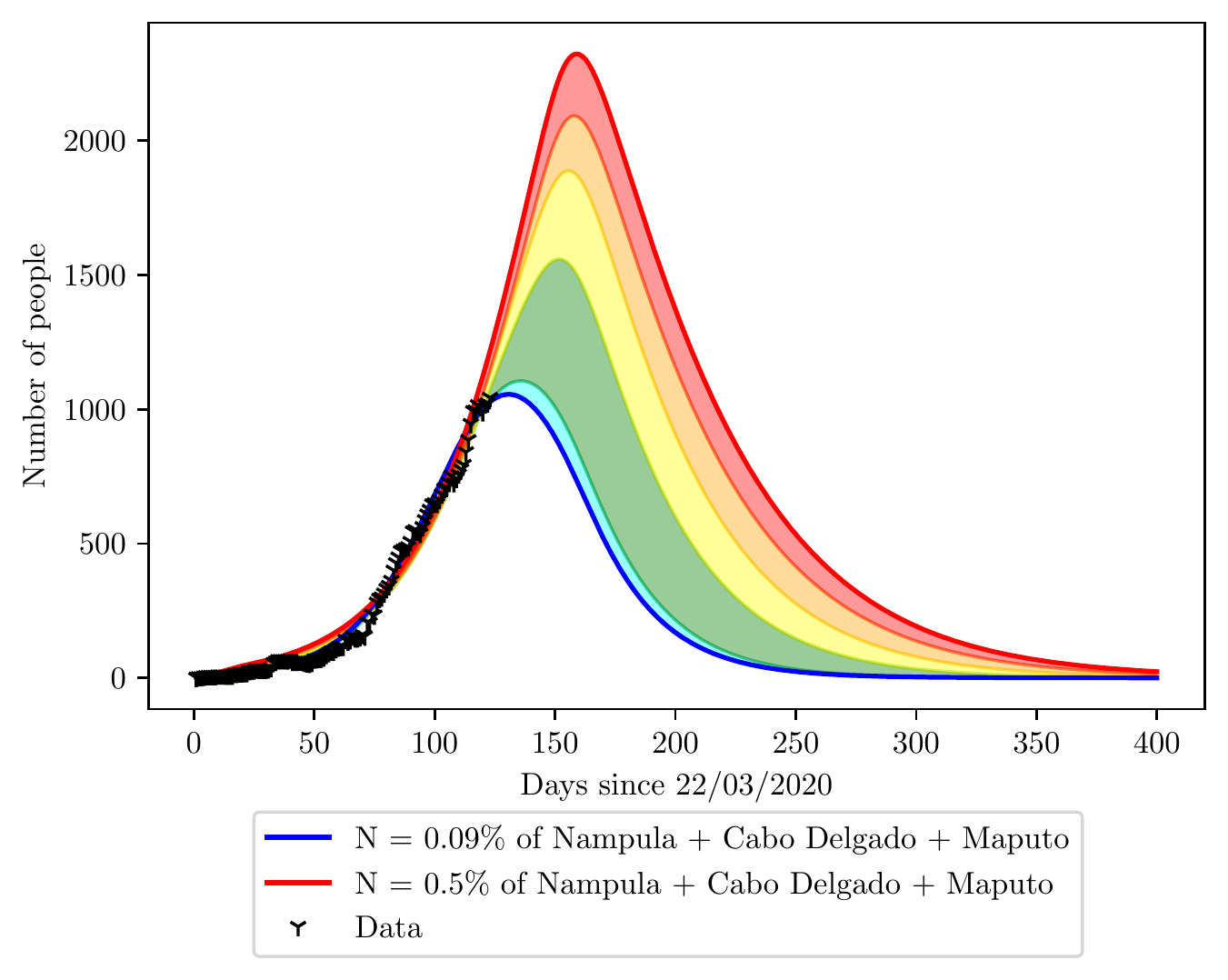}
    \caption{Predictions for the future behavior of the outbreak considering a lockdown with indefinite duration starting at the beginning of August, that is, close to the 130th day after the first case.}
    \label{fig.Lockdown}
  \end{minipage}
  \begin{minipage}[b]{0.49\textwidth}
    \includegraphics[width=\textwidth]{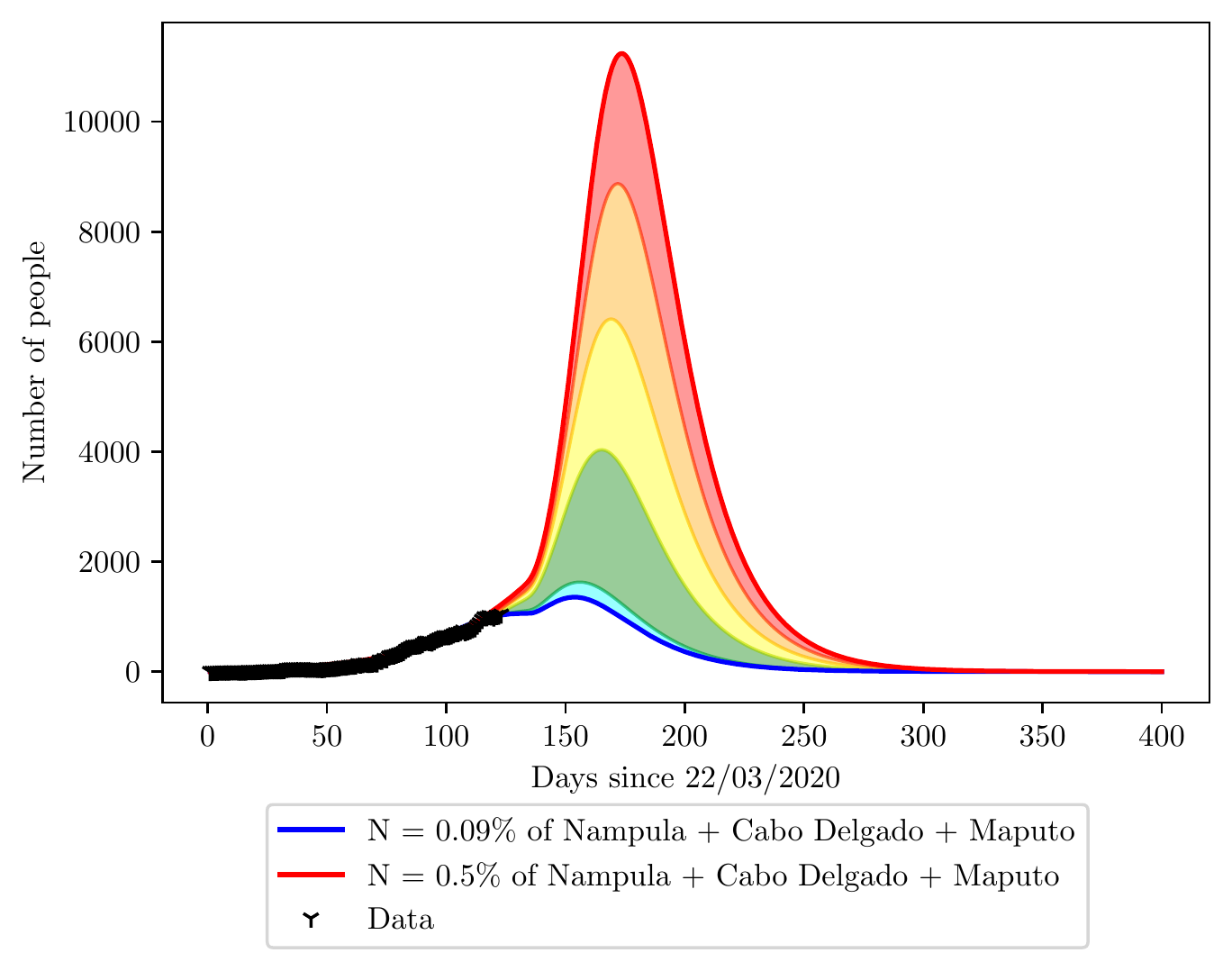}
    \caption{Predictions for the future behavior of the outbreak considering a re-opening of schools and commerce at the beginning of August, close to the 130th day after the first case.}
    \label{fig.ReopenFast}
  \end{minipage}
  \hfill
  \begin{minipage}[b]{0.49\textwidth}
    \includegraphics[width=\textwidth]{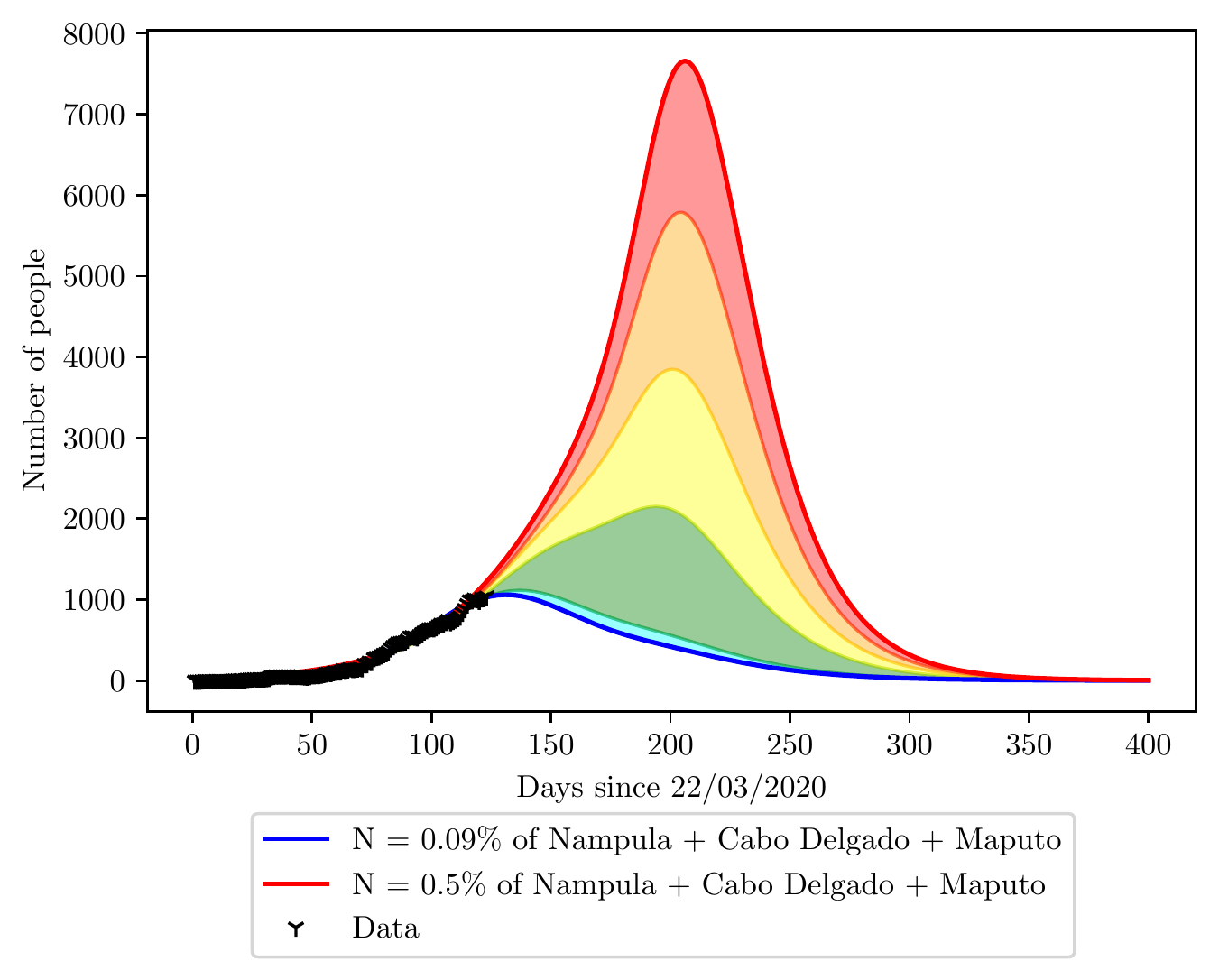}
    \caption{Predictions for the future behavior of the outbreak considering a 10 times slower re-opening of schools and commerce at the beginning of August, close to the 130th day after the first case.}
    \label{fig.ReopenSlow}
  \end{minipage}
\end{figure}

After the adjustment, we considered some other possible scenarios for the future behavior of the outbreak. The first was considering a national lockdown taking place at the beginning of August (Figure \ref{fig.Lockdown}), the second was simulating a full re-opening of commerce, schools and daily life return in 10 days after the first day of August (Figure \ref{fig.ReopenFast}), the third scenario was a slower re-opening, 10 times slower than the previous case (Figure \ref{fig.ReopenSlow}). The aim when considering these projections is to provide useful information for policy management when thinking about the future dynamics of the pandemic crisis.

According to these projections the return to daily activities on August could increase the total number of COVID-19 infections at the peak 3.4 times, depending on the effective $N$, ranging from a 1.2 increase in peak height to 3.4. By the other hand, a slower return, according to the consideration in which the infection rate $\beta$ increases 10 times slower, could provoke an increase of the peak number from 0 to 2.3. Finally, the lockdown may decrease the infection peak to 60\% of the current maximum tendency, which is represented by the red curve in figure \ref{fig.Lockdown}. At the best scenario, represented by the blue and cyan curves in figure \ref{fig.Lockdown}, we see that the lockdown would not affect the peak height, however it would result in a faster decrease of the infection curve after the peak.

\begin{figure}[ht]
\centering
\begin{subfigure}{.4\textwidth}
  \centering
  \includegraphics[width=1\linewidth]{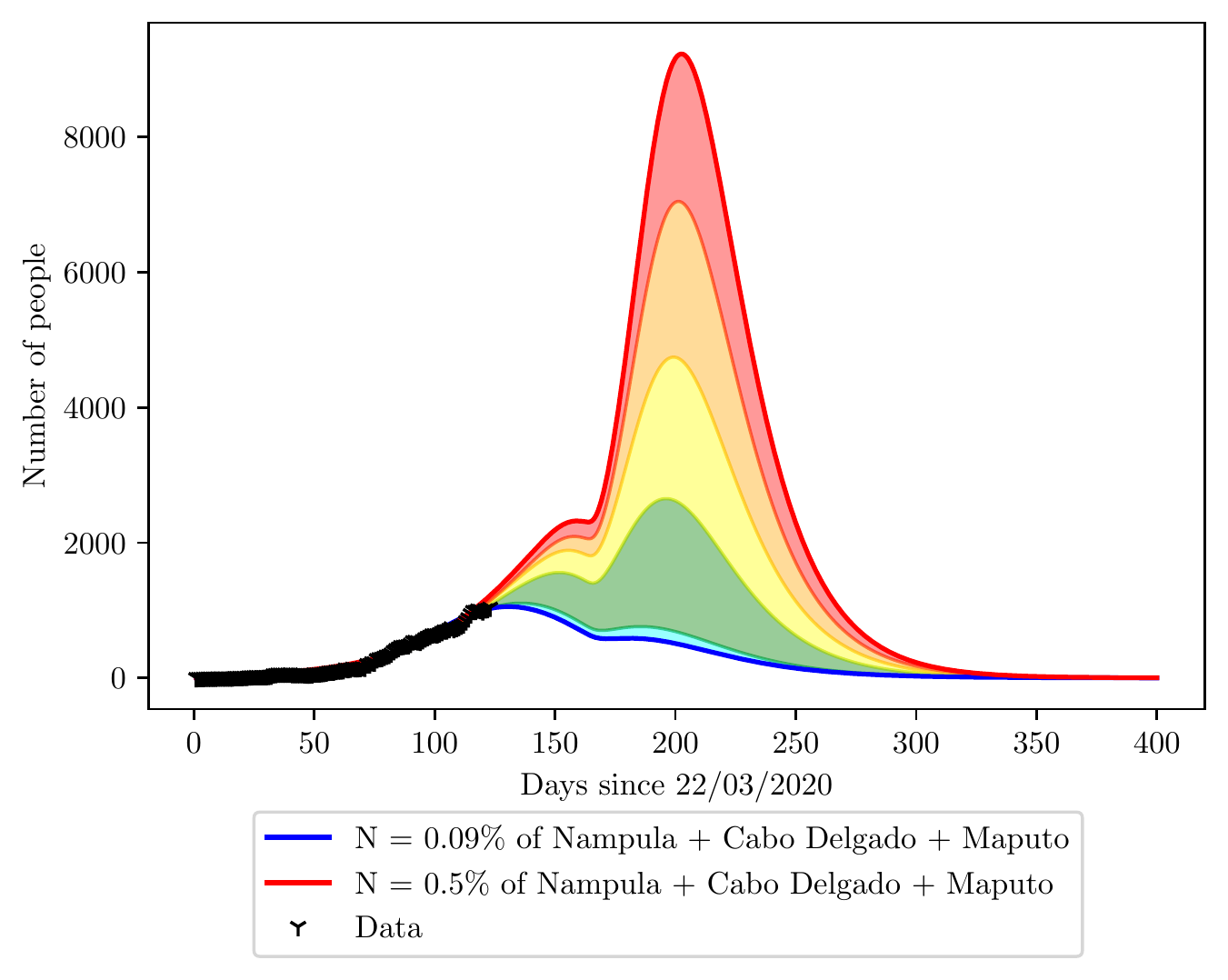}  
  \caption{Predictions considering a lockdown of 30 days with a fast return to daily life.}
  \label{fig.Lockdown30fast}
\end{subfigure}
\begin{subfigure}{.4\textwidth}
  \centering
  \includegraphics[width=1\linewidth]{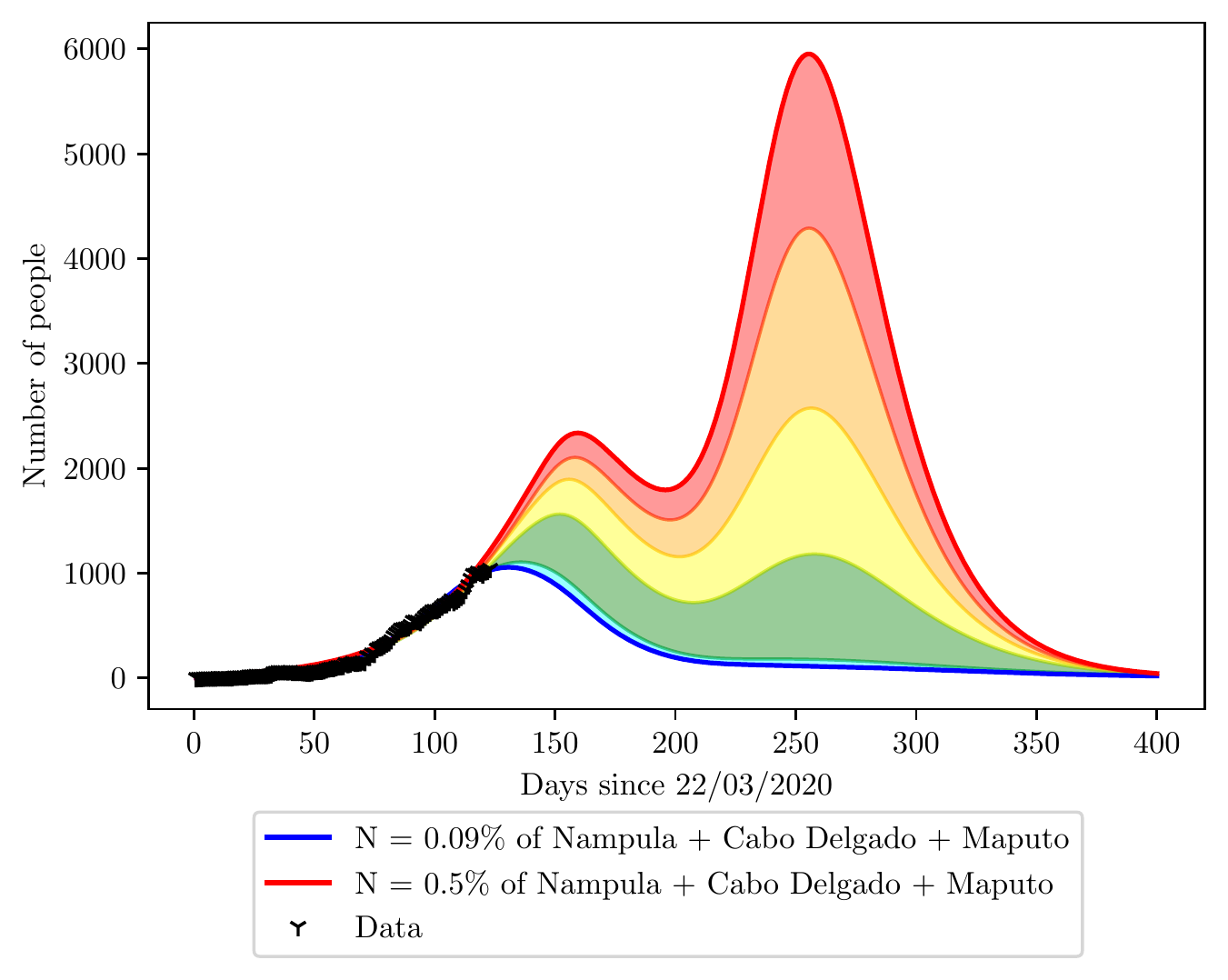}  
  \caption{Predictions considering a lockdown of 30 days with a slow return to daily life.}
  \label{fig.Lockdown30slow}
\end{subfigure}

\begin{subfigure}{.4\textwidth}
  \centering
  \includegraphics[width=1\linewidth]{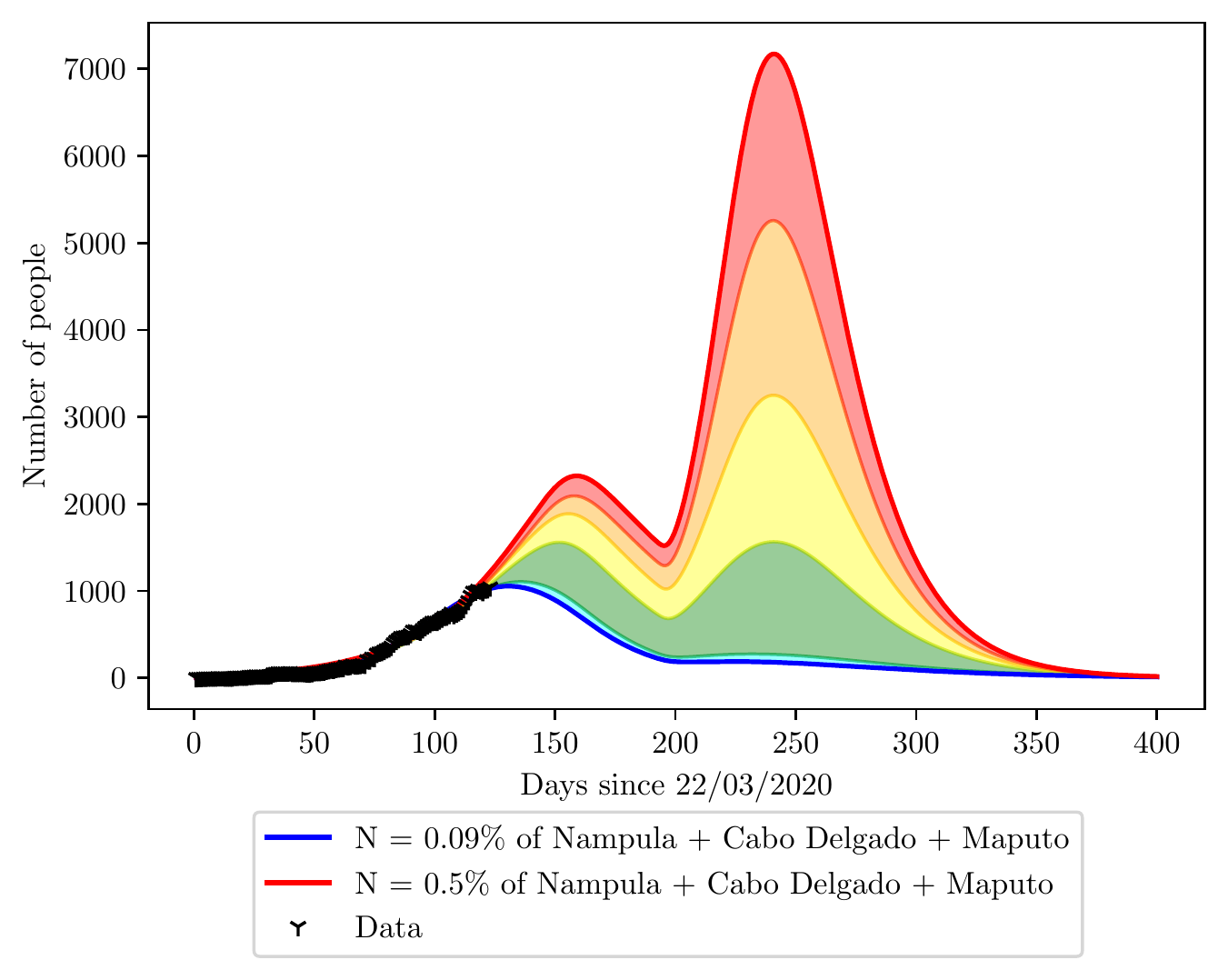}  
  \caption{Predictions considering a lockdown of 60 days with a fast return to daily life.}
  \label{fig.Lockdown60fast}
\end{subfigure}
\begin{subfigure}{.4\textwidth}
  \centering
  \includegraphics[width=1\linewidth]{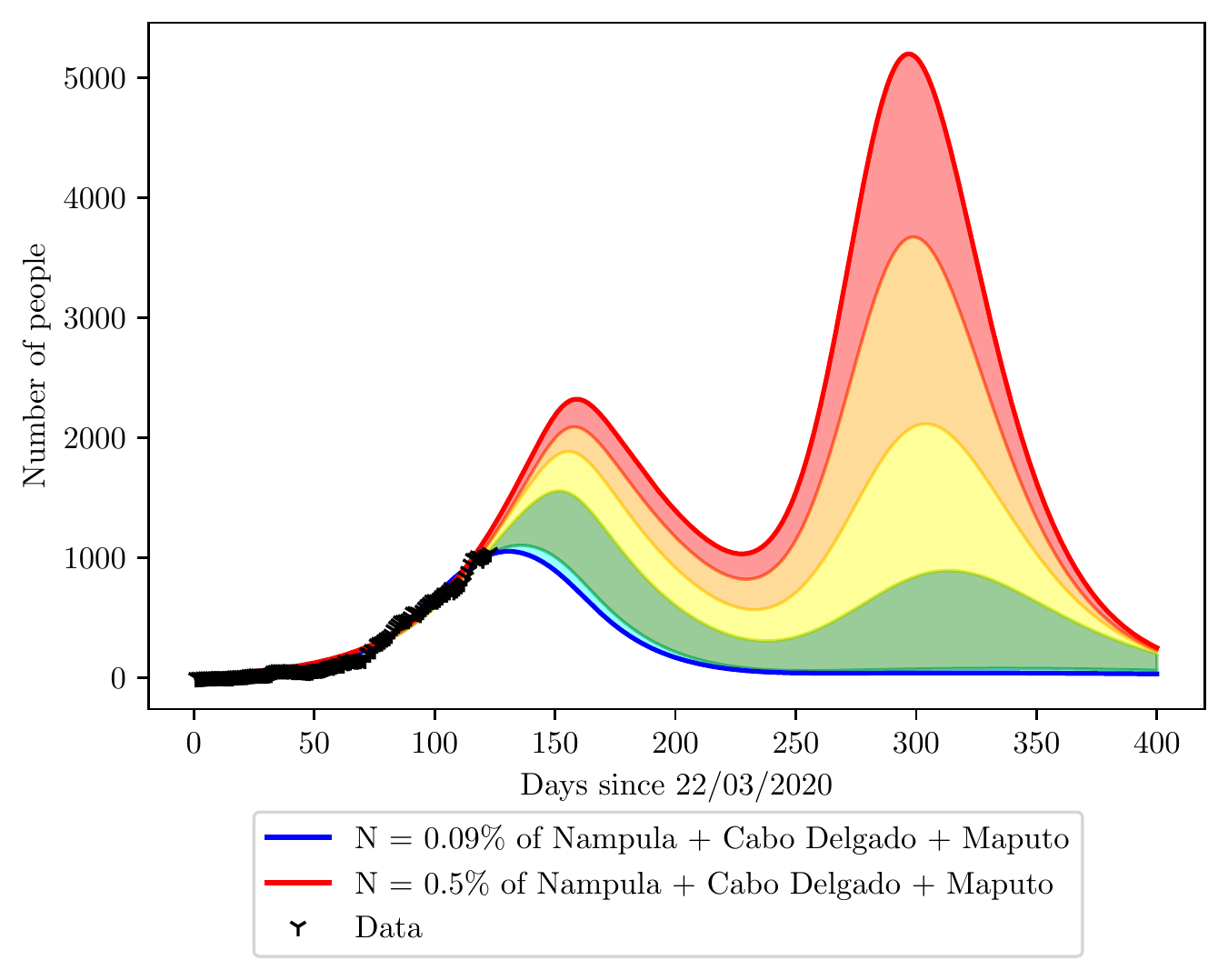}  
  \caption{Predictions considering a lockdown of 60 days with a slow return to daily life.}
  \label{fig.Lockdown60slow}
\end{subfigure}

\begin{subfigure}{.4\textwidth}
  \centering
  \includegraphics[width=1\linewidth]{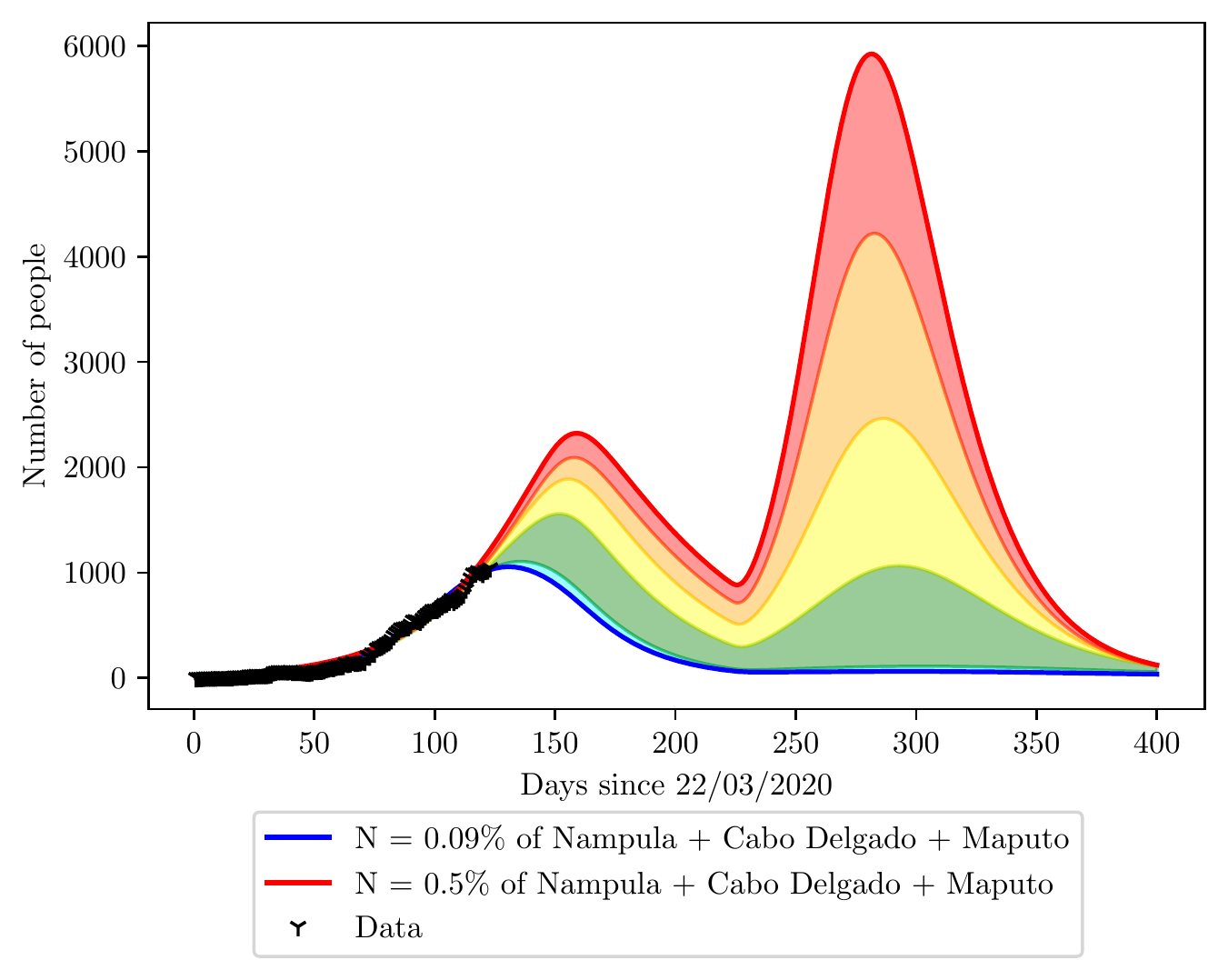}  
  \caption{Predictions considering a lockdown of 90 days with a fast return to daily life.}
  \label{fig.Lockdown90fast}
\end{subfigure}
\begin{subfigure}{.4\textwidth}
  \centering
  \includegraphics[width=1\linewidth]{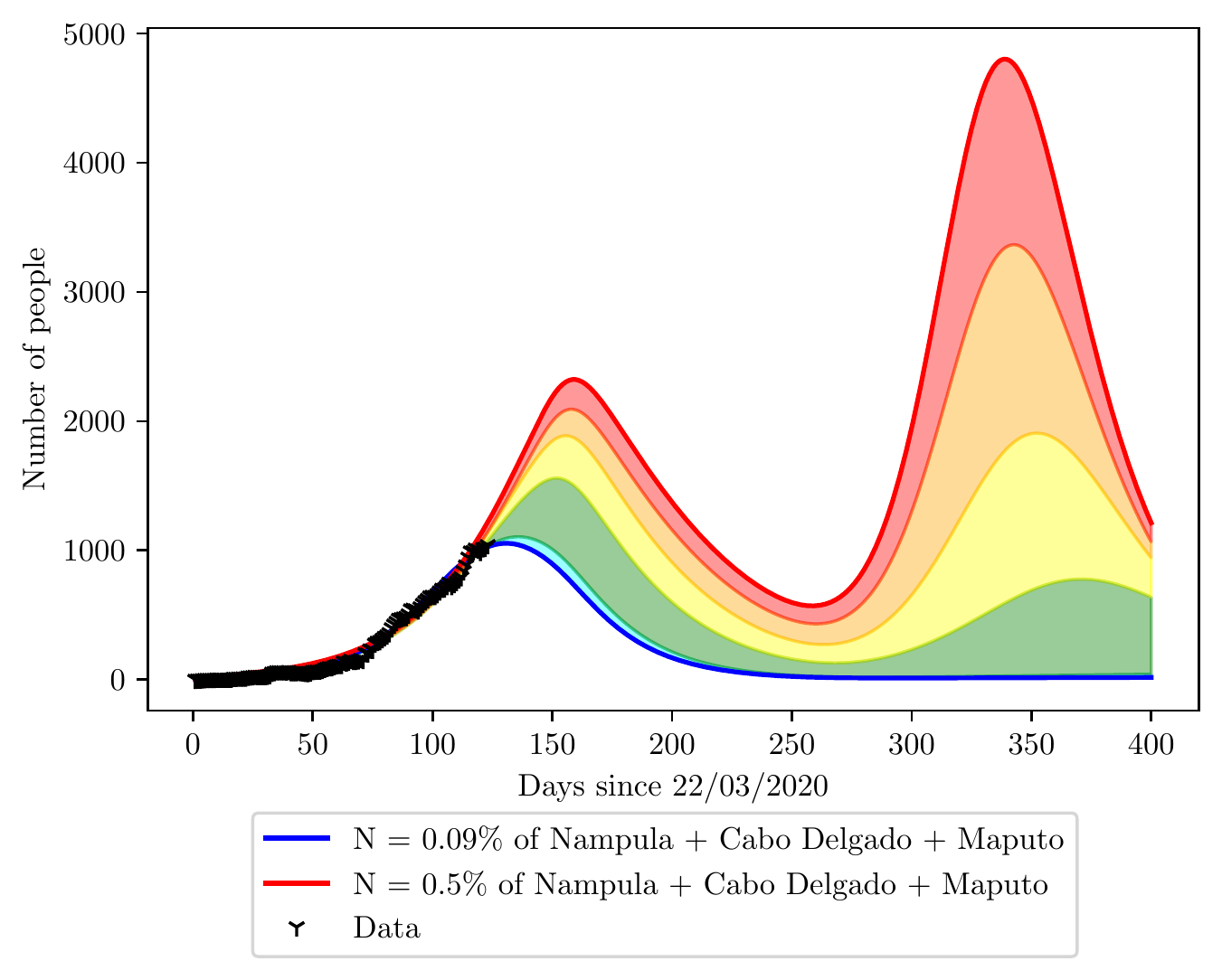}  
  \caption{Predictions considering a lockdown of 90 days with a slow return to daily life.}
  \label{fig.Lockdown90slow}
\end{subfigure}
\caption{The color gradient represents each $N$ simulated, blue curve is the simulated scenario with $N = 0.09\%$, cyan region represents $N=0.1\%$, green $N=0.2\%$, yellow = $N=0.3\%$ and so on. Figures \ref{fig.Lockdown30fast} and \ref{fig.Lockdown30slow} represent the behavior for a lockdown lasting 30 days, taking place at the beggining of August, at the 130th day of pandemic. They both represent the scenarios where the return to normal activies in commerce, schools and home isolation, is taken fast (\ref{fig.Lockdown30fast}) and slow (\ref{fig.Lockdown30slow}). Figures \ref{fig.Lockdown60fast} and \ref{fig.Lockdown60slow} represent the same scenario, however with a longer duration for the lockdown; 60 days. Finally, figures \ref{fig.Lockdown90fast} and \ref{fig.Lockdown90slow} shows the final scenario where the lockdown lasts 90 days.}
\label{fig.lockdowns}
\end{figure}

As expected, a national lockdown cannot be held forever until the last infection registered due to economical costs. We then, estimated 6 scenarios for duration of a lockdown and re-opening; corresponding to a lockdown for 30, 60 and 90 days and the same situation considering 30, 60 and 90 days of lockdown with a slow return to daily activities (Figures \ref{fig.Lockdown30fast} to \ref{fig.Lockdown90slow}).

The projections show that a slower re-opening might decrease the second peak height by 36\%, considering a lockdown that lasts 30 days, 29\%, for 60 days duration, and 20\% for a 90 days duration. On average, the gradual return to daily life shows the potential to decrease the second wave intensity by 28\%.

Considering the duration of the lockdown, the longer the duration time for the intervention, the lower the second peak is after return.

\section{Discussion}
\label{sec.Discussion}

Our aim in this study is to evaluate the future dynamics of COVID-19 pandemic crisis in Mozambique. We have shown that simulations predict over 200\% increase of peak intensity if commercial activities and daily life returns to normal in August, when compared to the tendency of cases. Moreover, lockdown policies and slower returns substantially decrease the peak intensity.

The main limitation of our study is the effective population $N$ susceptible to the disease. In the real situation, $N$ increases with time, as the disease progresses and the virus begging circulating in more interior regions. That behavior could provoke an increase in the susceptible population, causing more abrupt peaks and creating a bigger number of cumulative cases. Therefore, our estimations here should not be taken as a precise prediction regarding the numbers of infections, instead, we draw different scenarios and show how different policies might change the infection curve, given the recent data for Mozambique.

Another limitation lies on the data, which might be biased by a lower testing rate. However, to date, Mozambique performed 31.7 tests for each positive case, on average, which is close to Spanish numbers according to the online platform Ourworldindata (\url{https://ourworldindata.org/coronavirus-testing}). In countries known to have acquired a accurate tracking of cases, such as Germany, South Korea and Australia, these numbers jump to 191.9, 201.5 and 196.8, respectively. By the other hand, countries known to have a poor accuracy in tracking of cases such as Brazil and United States have 1.5 and 11.5 tests per confirmed case, respectively. We could estimated that Mozambique data is not perfect and might lose some cases, provoking changes on the future of the pandemic, however it is not poorly tracked.

The last limitation lies in the model and the algorithm itself; first, the model does not take into account infections caused by contact infected surfaces, which is known to be a viable source of infection due to viral persistence at inanimate surfaces \cite{kampf2020persistence}. Second, the algorithm used is simple and direct, but is susceptible to the nonidentifiability problem \cite{roda2020difficult}, which arises in non statistical fitting methods. However this is not expected to change drastically the shape of the curve.

Our study also found a positive growth rate, indication of a continuous growth of cases in Mozambique. The $\mathcal{R}_t$ number is however low, suggesting a slow increase of infection curve. However, due to the lower number of infections, to date, Mozambique might still be on a stochastic limit of behavior, making the measurement of $\mathcal{R}_t$ imprecise, but not very different from the value found here.

The infection curve might be close to the infection peak, considering the best scenario, however it is also possible to observe a prolonged increase, as the worse scenario curves have shown. It is not possible to say which curve the country will tend to follow, this mainly depends on the effective population in reach of the disease. Therefore, we highlight that our study represents simple views of the real scenario and should not represent the precise future of Mozambique, instead it should provide mere estimates for the possible behaviors.

\section*{Conflict of Interest}

The authors declare no conflict of interest.

\section*{Funding}

The study received no external funding or resources.

\section*{Acknowledgments}

Our thanks to Francesse Mauro and Márcio Mathe, from the Department of Mathematics, course of Geographical Information System, in the Universidade Eduardo Mondlane, to providing us the map for the demographic density of Mozambique.

Cláudio Moisés Paulo thanks to the project Development in Africa with Radio Astronomy (DARA, \url{https://www.dara-project.org/}) and to the project Development of Palop Knowledge in Radioastronomy (DOPPLER\footnote{DOPPLER is funded by the Aga Khan Development Network and the Fundação para a Ciência e a Tecnologia (FCT), under grant number 333197717.}, \url{http://doppler.av.it.pt/}), for providing him equipment which was used for this study.

\section*{Additional Information}

Our source code is freely available at \url{https://github.com/PedroHPCintra/COVID-Mozambique}.

\bibliographystyle{unsrt}  
\bibliography{references}  


\end{document}